# Building a physics-aware AI ecosystem for solid-state hydrogen storage materials

*Seong-Hoon Jang[1, 2], Yiwen Yao[3], Chuanyu Liu[4], Linda Zhang[*1,5], Di Zhang[1,5], Xue Jia[1], Hung Ba Tran[1], Eric Jianfeng Cheng[1], Ryuhei Sato[6], Yusuke Ohashi[7], Toyoto Sato[7], Yusuke Hashimoto[1,5], Mark Allendorf[8], Nongnuch Artrith[9], Marcello Baricco[10], Andreas Borgschulte[1,11], Darren P. Broom[12], Ang Cao[13], Benjamin W. J. Chen[14], Lixin Chen[15], Ping Chen[16], Eun Seon Cho[17], Stefano Deledda[18], Zhao Ding[19], Martin Dornheim[20], Michael Felderhoff[21], Yaroslav Filinchuk[22], George E. Froudakis[23], Mingxia Gao[24], Thomas Gennett[25], Zaiping Guo[26], Ikutaro Hamada[27,28], Jason Hattrick-Simpers[29], Bjørn C. Hauback[18], Michael Hirscher[1,30], Torben R. Jensen[31], Baohua Jia[32], Hyoung Seop Kim[33], Takahiro Kondo[1,34,35], Kentaro Kutsukake[36], Xiao-Yan Li[3], Tongliang Liu[38], Piao Ma[39,40], Jianfeng Mao[41], Rana Mohtadi[42], Hyunchul Oh[43], Mark Paskevicius[44], Chris J. Pickard[1,45], Astrid Pundt[46], Anibal Ramirez-Cuesta[47], Hiroyuki Saitoh[48], Kaihang Shi[49], Aloysius Soon[50], Chenghua Sun[51], Chris Wolverton[52], Hiroshi Yabu[1], Weijie Yang[53], Zhenpeng Yao[54,55], Xuebin Yu[56], Jianxin Zou[54], Shouyi Hu[54], Panpan Zhou[24], Xi Lin[54], Zhigang Hu[54,55], Zhenhao Zhou[54,55], Pengfei Ou[*3], Jiayu Peng[*4], Shin-ichi Orimo[*1, 7], and Hao Li[*1]*

[1] Advanced Institute for Materials Research (WPI-AIMR), Tohoku University, Sendai 980-8577, Japan

[2] Unprecedented-scale Data Analytics Center, Tohoku University, Sendai 980-8578, Japan

[3] Department of Chemistry, National University of Singapore, Singapore 117549

[4] Department of Materials Design and Innovation, University at Buffalo, Buffalo, NY 14260, USA

[5] Frontier Research Institute for Interdisciplinary Sciences (FRIS), Tohoku University, Sendai 980-0845, Japan

[6] Department of Materials Engineering, The University of Tokyo, Tokyo 113-8656, Japan

[7] Institute for Materials Research, Tohoku University, Sendai, 980-8577, Japan

[8] Department of Chemistry, Washington University in St. Louis, St. Louis, MO 63130, USA

[9] Debye Institute for Nanomaterials Science, Utrecht University, 3584 CC Utrecht, Netherlands

[10] Department of Chemistry and NIS, University of Turin, Turin 10125, Italy

[11] Empa Swiss Federal Laboratories for Materials Science and Technology, Chemical Energy Carriers and Vehicle Systems Laboratory, Dübendorf, 8600, Switzerland
[12] Hiden Isochema Ltd, Warrington WA5 7UN, UK
[13] State Key Laboratory of Clean Energy Utilization, College of Energy Engineering, Zhejiang University, Hangzhou 310027, China
[14] Institute of High Performance Computing, Agency for Science, Technology and Research, Singapore 138632, Singapore
[15] School of Materials Science and Engineering, Zhejiang University, Hangzhou 310027, China
[16] Dalian Institute of Chemical Physics, Chinese Academy of Sciences, Dalian 116023, China
[17] Department of Chemical and Biomolecular Engineering, Korea Advanced Institute of Science and Technology (KAIST), Daejeon 34141, Republic of Korea
[18] Department for Hydrogen Technology, Institute for Energy Technology (IFE), 2007 Kjeller, Norway
[19] National Engineering Research Center for Magnesium Alloys, College of Materials Science and Engineering, Chongqing University, Chongqing 400044, China
[20] Advanced Materials Research Group, Faculty of Engineering, University of Nottingham, Nottingham NG7 2RD, UK
[21] Max-Planck-Institut für Kohlenforschung, Mülheim an der Ruhr 45470, Germany
[22] Institute of Condensed Matter and Nanosciences (IMCN), UCLouvain, Louvain-la-Neuve 1348, Belgium
[23] Department of Chemistry, University of Crete, Voutes Campus, Heraklion GR-70013, Greece
[24] State Key Laboratory of Silicon and Advanced Semiconductor Materials, School of Materials Science and Engineering, Zhejiang University, Hangzhou 310058, Zhejiang, China
[25] National Renewable Energy Laboratory (NREL), Golden, CO 80401, USA
[26] Department of Materials Science and Engineering, City University of Hong Kong, Kowloon, Hong Kong SAR 999077, China
[27] Department of Precision Engineering, Graduate School of Engineering, The University of Osaka, 2-1 Yamadaoka, Suita, Osaka 565-0871, Japan
[28] Research Center for Precision Engineering, Graduate School of Engineering, The University of Osaka, 2-1 Yamadaoka, Suita, Osaka, 565-0871, Japan

[29] Department of Materials Science & Engineering, University of Toronto / CanmetMATERIALS, Toronto, ON M5S 3E4, Canada
[30] Max Planck Institute for Solid State Research, 70569 Stuttgart, Germany
[31] iNANO & Department of Chemistry, Aarhus University, DK-8000 Aarhus, Denmark
[32] Centre for Atomaterials and Nanomanufacturing, RMIT University, Melbourne, VIC 3001, Australia
[33] Graduate Institute of Ferrous Technology (GIFT), Pohang University of Science and Technology (POSTECH), Pohang 37673, Republic of Korea
[34] Department of Materials Science, Faculty of Pure and Applied Sciences, University of Tsukuba, Tsukuba 305-8573, Japan
[35] Hydrogen Boride Research Center, Tsukuba Institute of Advanced Research, University of Tsukuba, Tsukuba 305-8577, Japan
[36] Institute of Materials and Systems for Sustainability (IMaSS), Nagoya University, Nagoya 464-8603, Japan
[37] Department of Chemistry, National University of Singapore (NUS), Singapore 117549, Singapore
[38] School of Computer Science, Faculty of Engineering, University of Sydney, Sydney, NSW 2006, Australia
[39] Suzhou MatSource Technology Co., Ltd., Suzhou 215000, Jiangsu, China
[40] Gusu Laboratory of Materials, Suzhou 215000, Jiangsu, China
[41] School of Chemical Engineering, University of Adelaide, Adelaide, SA 5005, Australia
[42] Toyota Research Institute of North America (TRI-NA), Ann Arbor, MI 48105, USA
[43] Department of Chemistry, Ulsan National Institute of Science and Technology (UNIST), Ulsan 44919, Republic of Korea
[44] Department of Physics and Astronomy, Institute for Energy Transitions, Curtin University, Perth, WA 6845, Australia
[45] Department of Materials Science & Metallurgy, University of Cambridge, Cambridge CB3 0FS, UK
[46] Institute of Applied Materials (IAM-WK), Karlsruhe Institute of Technology (KIT), Karlsruhe 76131, Germany

[47] Neutron Technologies Division, Oak Ridge National Laboratory (ORNL), Oak Ridge, TN 37831, USA

[48] Kansai Institutes for Photon Science, National Institutes for Quantum Science and Technology (QST), Hyogo 679-5148, Japan

[49] Department of Chemical and Biological Engineering, University at Buffalo, The State University of New York, Buffalo, New York 14260, United States

[50] Department of Materials Science and Engineering, Yonsei University, Seoul 03722, Republic of Korea

[51] Computational Materials and Catalysis Group, School of Science, Swinburne University of Technology, Melbourne, VIC 3122, Australia

[52] Department of Materials Science and Engineering, Northwestern University, Evanston, IL 60208, USA

[53] Department of Power Engineering, North China Electric Power University, Baoding, Hebei 071003, China

[54] Shanghai Key Laboratory of Hydrogen Science & Center of Hydrogen Science, School of Materials Science and Engineering, Shanghai Jiao Tong University, Shanghai 200240, China

[55] Innovation Center for Future Materials, Zhangjiang Institute for Advanced Study, Shanghai Jiao Tong University, Shanghai 201203, China

[56] Department of Materials Science, Fudan University, Shanghai 200433, China

Corresponding authors:

linda.zhang.a3@tohoku.ac.jp (L. Zhang)

pengf.ou@nus.edu.sg (P. Ou)

jypeng@buffalo.edu (J. Peng)

shin-ichi.orimo.a6@tohoku.ac.jp (S. Orimo)

li.hao.b8@tohoku.ac.jp (H. Li)

ABSTRACT

Hydrogen storage remains a central bottleneck for scalable hydrogen energy systems due to the multiscale and coupled nature of the thermodynamics, kinetics, and microstructural evolution of hydrogen storage materials (HSMs). Although artificial intelligence (AI) has accelerated materials discovery, current approaches remain constrained by fragmented data, limited physical consistency, and weak integration with experimental validation. Here, we propose a unified framework that integrates coherent data infrastructure, physics-grounded modeling, and AI-driven inverse design within a closed-loop discovery paradigm. By embedding physical constraints and experimental feedback, this approach enables adaptive, physically consistent optimization, thereby establishing a pathway toward autonomous, digital-twin-enabled discovery of HSMs.

KEYWORDS. Hydrogen storage materials, database, machine learning interatomic potentials, generative models, large language models, closed-loop discovery, digital twin.

## Introduction

Hydrogen is widely regarded as a cornerstone of future carbon-neutral energy systems, yet its deployment remains constrained by the lack of efficient, safe, low-cost, and reversible storage solutions.[1,2] Its low density under ambient conditions (0.082 kg·m$^{-3}$, corresponding to a volumetric energy density below 10 MJ·m$^{-3}$) necessitates densification strategies for practical use. While compression (35-70 MPa) and liquefaction (<25 K) are established approaches, both impose significant energy penalties and engineering challenges, limiting scalability. These limitations originate from the intrinsically weak intermolecular interactions of dihydrogen, which hinder dense packing. Here, we focus specifically on solid-state hydrogen storage materials (HSMs), thereby excluding liquid organic hydrogen carriers. Within this framework, two broad strategies have therefore emerged: *physisorption* in high-surface-area porous solids and *chemisorption via* metal–hydrogen bond formation, each with distinct advantages and limitations.[3]

Physisorption-based storage relies on weak gas-surface interactions and is most effective at low temperatures. Nanoporous materials such as metal-organic frameworks (MOFs), covalent organic frameworks (COFs), hydrogen-bonded organic frameworks (HOFs), and porous organic polymers (POPs) offer high surface areas and tunable pore environments, enabling systematic exploration of structure–property relationships.[4-7] However, low adsorption enthalpies limit performance near ambient conditions, motivating efforts to enhance binding strength without sacrificing reversibility.

In contrast, solid-state hydrogen storage *via* chemisorption in metal and complex hydrides offers significantly higher volumetric densities.[8-10] In these systems, hydrogen uptake and release involve bond formation and phase transformations, leading to favorable storage capacities but also introducing challenges related to thermodynamics and kinetics. The behavior of such materials is commonly characterized by pressure-composition isotherms, from which key quantities such as enthalpy, entropy, and maximum storage capacity can be extracted. In practice, however, reliable datasets remain scarce due to sluggish kinetics, contamination effects, and sensitivity to microstructural and processing conditions, which hinder both fundamental understanding and the rational design of materials and systems.

Beyond these two archetypes, emerging material classes (including compositionally complex high-entropy alloys) offer additional opportunities to tune hydrogen interactions through multicomponent design.[11] Despite this diversity, the discovery of high-performance HSMs remains slow and largely empirical, reflecting the intrinsic complexity of coupled thermodynamic and kinetic processes. In addition, HSMs span distinct chemical classes (i.e., including metal hydrides, complex hydrides, and nanoporous frameworks) that often exhibit limited transferability of design principles, effectively forming "islands" of chemistry. Also, additional classes (i.e., non-reversible chemical hydrides, hydrolysis-based systems, and liquid organic hydrogen carriers) present distinct challenges and design criteria, further broadening the hydrogen storage landscape. **Hydrogen storage performance is governed not only by capacity but also by equilibrium pressure, operating temperature, diffusion kinetics, phase transformations, and long-term cycling stability, reflecting a fundamentally multi-objective optimization problem.**[12] Crucially, these properties are path-dependent, evolving across multiple length and time scales and remaining highly sensitive to activation history, defect structure, impurities, and operating conditions. In addition, system-level factors (e.g., heat and mass transport and tank design) impose further constraints on practical performance, collectively limiting predictive design.[13]

Recent advances in data-driven materials science and artificial intelligence (AI) offer new opportunities to address these challenges.[14-20] High-throughput computational screening, curated experimental datasets, and machine learning (ML) models enable rapid exploration of vast design spaces and identification of promising candidates. However, progress is currently constrained by fragmentation across data sources and by the limited integration of physically meaningful descriptors, kinetic effects, and degradation processes. As a result, model robustness and transferability remain limited.

Addressing these challenges requires a more tightly integrated framework that connects data generation, model development, and experimental validation. While automation and self-driving laboratory (SDL) approaches can accelerate data acquisition and improve reproducibility, their primary value lies in enabling the generation of consistent, high-quality datasets that capture condition-dependent behavior. When coupled with physics-informed models, such data can support more reliable predictions and iterative refinement of materials design strategies, rather than serving as an end in themselves.

**A new paradigm is therefore emerging that integrates data infrastructure, physics-grounded modeling, and adaptive experimentation within a unified, closed-loop framework.** In this context, hydrogen storage remains inherently multiscale and path-dependent, requiring approaches that explicitly account for thermodynamic consistency and reproducibility. The current capabilities,

limitations, and open challenges are summarized in **Table 1**, highlighting key bottlenecks and opportunities for integration across these domains.

Here, based on the current state of the field and in-depth discussions among 69 active researchers from the global communities of HSMs, AI, SDLs, and computational science, we outline a pathway toward an integrated, physics-aware ecosystem for solid-state HSM discovery. By combining structured data, interpretable models, and iterative experimental validation, this approach aims to accelerate the identification and optimization of HSMs, ultimately enabling scalable and efficient energy systems.

**Table 1. Maturity and key challenges of AI-driven components in hydrogen storage materials discovery.**

| Component | Current capabilities | Key limitations | Next challenges |
|---|---|---|---|
| Data infrastructure | Curated datasets, high-throughput computation, high-throughput experiments | Fragmentation, datasets with incompatible formats and schemas, inconsistent datapoints across datasets, missing kinetics/degradation, incomplete experimental context, limited reproducibility assessment | Interoperable, multimodal, thermodynamically consistent datasets, and reproducibility-aware datasets with uncertainty, provenance tracking, data augmentation strategies |
| Physics-grounded modeling with machine learning interatomic potentials (MLIPs) | Atomistic dynamics, diffusion, defects, phase transformation | Limited transferability, free-energy accuracy | Integration with thermodynamics and multiscale models, open-system formalisms, and multiscale models under experimentally relevant conditions |
| Generative models (GMs) | Structural exploration, inverse design | Lack of physical constraints, uncertain synthesizability, limited experimental grounding, out of distribution predictions | Physics-constrained generation and coupled with thermodynamic validation and synthesis-aware screening |
| Large language models (LLMs) | Knowledge extraction, hypothesis generation, multimodal literature mining | Lack of physical grounding, hallucinations, alignment, loss of context, variable | Integration with physical models and databases, provenance-aware reasoning frameworks |
| | | reliability, limited treatment of uncertainty | |
| Active learning and uncertainty quantification | Active learning for computations, Bayesian optimization for experimentation | Separate acquisition frameworks for experiments and computations, different data fidelities | Hybrid, cost-aware experimental-computational acquisition strategies, |
| Closed-loop discovery | ML-guided experiments, early automation | Limited adaptability, integration gaps, weak uncertainty handling, insufficient data-quality awareness | Autonomous, adaptive experimental systems, linked to reproducibility-aware databases and confidence-calibrated decision-making |

**Data Infrastructure for Hydrogen Storage**

The central bottleneck in data-driven hydrogen storage research is no longer the lack of predictive algorithms, but the absence of a thermodynamically coherent and interoperable data infrastructure grounded in experiment. Hydrogen storage data exist in two partially disconnected layers (**Figure 1a**): (i) computational repositories containing density functional theory (DFT) energies, structures, and descriptors, and (ii) heterogeneous experimental reports, where key information is often buried in figures or narrative text, spread across formats and languages, and difficult to retrieve with standard search methods. This fragmentation forces reliance on expert interpretation and limits reliable ML and inverse design.

Computational repositories have advanced significantly. Databases such as the Materials Project, OQMD, AFLOW, NOMAD, and Alexandria provide large-scale access to computed structures and energetics, enabling high-throughput screening.[21-25] However, these datasets are biased toward equilibrium properties under idealized conditions and generally lack surface and interface information. For solid-state HSMs, where performance depends on pressure-temperature behavior, kinetics, and cycling stability, static computational data provide an incomplete representation. Moreover, the lack of explicit linkage between computed entries and experimentally validated systems limits their applicability.

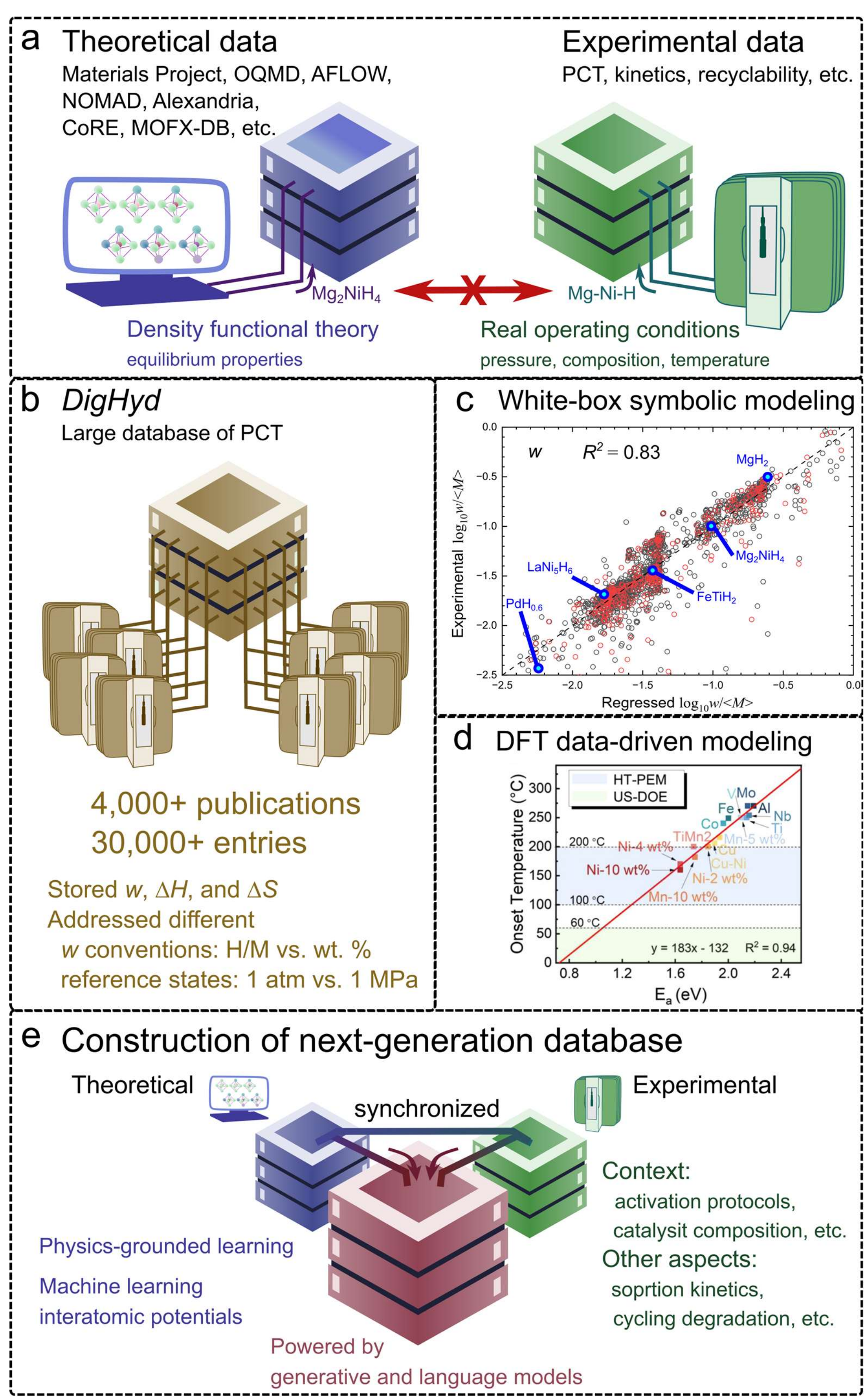

**Figure 1. Data infrastructure for solid-state hydrogen storage materials. (a)** Schematic comparison between computational and experimental data. Computational databases (e.g.,

Materials Project, OQMD, AFLOW, NOMAD, Alexandria, CoRE, QMOF, and MOFX-DB)[21-28] provide density functional theory (DFT)-derived equilibrium properties under idealized conditions, whereas experimental data reflect real operating environments, including pressure, temperature, compositional variability, impurities, surface conditions, measurement uncertainty, and bulk morphology. The lack of direct correspondence between these domains highlights a fundamental gap in current data infrastructure. **(b)** Representative example of curated experimental data from the Digital Hydrogen Platform (*DigHyd*), comprising > 4,000 publications and >30,000 entries of thermodynamic measurements (e.g., hydrogen capacity $w$, enthalpy $\Delta H$, and entropy $\Delta S$).[34] Standardization across differing reporting conventions (e.g., H/M vs wt. % and reference states) was addressed. **(c)** *DigHyd* enables the application of white-box symbolic modeling (as an example, a symbolic modeling case of gravimetric storage density ($w$) is represented), revealing physically interpretable relationships between composition and hydrogen storage properties.[39-41] Reproduced from Ref. 39, under the terms of the Creative Commons CC BY-NC license. **(d)** Data-driven regression linking dehydrogenation barrier ($E_a$) and onset temperature for $MgH_2$-based materials, highlighting the gap between current material performance and US Department of Energy (US-DOE) and the high-temperature proton exchange membrane (HT-PEM) targets.[42] Reproduced from Ref. 42, under the terms of the Creative Commons Attribution 4.0 International License (CC BY 4.0). **(e)** Conceptual framework for a next-generation hydrogen storage database integrating computational and experimental data. The unified platform incorporates thermodynamic, kinetic, and degradation-related and reproducibility-aware metadata, together with experimental context such as activation protocols, catalyst composition, gas purity, cycling history, and measurement provenance. Explicit treatment of uncertainty, negative/null results, and confidence-weighted data quality enables more reliable comparison between literature records and computational predictions. Physics-grounded learning and machine learning interatomic potentials enhance predictive consistency, while generative and large language models facilitate data integration, knowledge extraction and inverse design.

For nanoporous materials such as MOFs, high-throughput simulations have enabled large databases linking structure, pore characteristics, and adsorption behavior (e.g., CoRE, QMOF, and MOFX-DB).[26-28] These datasets more directly connect structure to performance under defined conditions, although still largely within idealized frameworks. Despite this progress, structure-property relationships remain limited by kinetic effects, phase transformations, and cycling complexity that are not fully captured in such models. More broadly, experimental and theoretical studies of MOFs and COFs have established key principles governing hydrogen adsorption.[4, 29] While pristine frameworks can achieve high gravimetric uptake under cryogenic conditions, their performance near ambient conditions is constrained by weak host-guest interactions. Efforts to enhance adsorption enthalpy (through alkali metal doping, transition metal functionalization, and electronic structure tuning) have demonstrated improved binding, but achieving practical storage performance under realistic conditions remains challenging.[30] Recent work combining high-throughput screening and ML is therefore shifting the focus toward rational design strategies that target both adsorption energetics and operating conditions.[31] Recent developments further incorporate data-driven approaches for synthesizability prediction and transfer learning across adsorption spaces in MOFs, enabling more efficient exploration of experimentally accessible materials.[32, 33]

Curated experimental databases have improved accessibility and consistency, partially bridging the gap between computational repositories and real systems. For example, the Digital Hydrogen Platform (*DigHyd*: [www.dighyd.org](www.dighyd.org)) extracts and standardizes hydrogen storage data from the literature (**Figure 1b**), aggregating thousands of publications and entries focused on thermodynamic quantities derived from pressure–composition–temperature measurements.[34] Such efforts address a key limitation: capacity alone is insufficient to evaluate performance, as thermodynamic and kinetic constraints ultimately determine usability. However, inconsistencies in reported quantities (e.g., hydrogen content definitions and reference states) continue to introduce uncertainty and complicate comparison. In addition, both computational and experimental datasets typically report single-point values without quantified uncertainties.

AI-assisted data extraction has enabled the construction of such datasets at scale. Frameworks such as the descriptive interpretation of visual expression (DIVE) approach enable systematic extraction of figure-based data, including pressure–composition isotherms and desorption profiles.[35] However, automated extraction introduces challenges related to accuracy, hallucination, and reproducibility, particularly when relying on large language models (LLMs). Retrieval-augmented approaches can partially mitigate these issues, but robust validation remains essential.

A more fundamental limitation lies in the absence of standardized data reporting protocols. Experimental hydrogen storage data are frequently presented in graphical form, resulting in loss of underlying measurement information and limiting reproducibility. This issue is particularly acute for microporous materials, where key adsorption features occur at very low relative pressures and are not easily extracted. Establishing machine-readable formats, analogous to the adsorption information file (AIF),[36] alongside standardized reporting guidelines for hydrogen sorption measurements,[37] would significantly improve data comparability and reuse. Furthermore, many hydrogen storage reactions involve amorphous or partially disordered intermediates that are not readily captured by periodic representations, highlighting the need for approaches that incorporate local structural descriptors or statistical models of disorder.

Scalability alone does not resolve the core challenge: hydrogen storage data are highly sensitive to experimental context, including activation history, processing history, microstructure, impurities, catalysts, surface and mechanical stress, and cycling conditions. Importantly, impurities and compositional inhomogeneities can lead to the formation of secondary phases, which fundamentally alter thermodynamic behavior rather than acting as minor perturbations. As a result, datasets lacking detailed contextual metadata can propagate errors into derived parameters and ML models.[38] Addressing this requires a shift toward reproducibility-aware data curation, where each record includes experimental provenance, measurement conditions, and confidence levels.

Such a framework should also distinguish between directly reported and extracted data, incorporate uncertainty quantification, and apply physics-based consistency checks across thermodynamic quantities. Importantly, negative and null results should be retained, as they define failure boundaries and improve model calibration. Together, these elements would transform hydrogen storage datasets from passive repositories into physically meaningful, confidence-weighted resources.

Despite progress, the current data landscape remains fragmented. While computational datasets provide structural descriptors and curated platforms organize thermodynamic data, other critical properties (e.g., kinetics, cycling degradation, hysteresis, activation barriers, impurity sensitivity, and phase evolution) remain sparse and inconsistently reported. This imbalance limits model robustness and transferability.

This fragmentation directly constrains predictive modeling. ML models trained on incomplete datasets may perform well within limited domains but fail to generalize across materials and conditions. In contrast, curated datasets with physically meaningful descriptors enable more reliable and interpretable models. For example, symbolic regression applied to curated hydrogen storage datasets has revealed compact, interpretable relationships between composition and thermodynamic properties (**Figure 1c**).[39-41] Similarly, data-driven analyses of Mg-based hydrides quantify dehydrogenation behavior and highlight gaps relative to application targets such as those defined by the US Department of Energy (**Figure 1d**).[42]

**A critical next step is the integration of experimental and computational data into a unified framework** (**Figure 1e**). On the experimental side, standardized identifiers, reproducibility-aware metadata, consistent reporting of key properties such as microstructure and phase constitution, and inclusion of negative results are required; the former are particularly important as microstructural features strongly influence kinetics. On the computational side, advances in physics-based learning and machine-learning interatomic potentials (MLIPs) offer improved representation of realistic behavior across time and length scales. Together, these developments point toward a hydrogen-storage-specific data ecosystem that is interoperable, physically grounded, and predictive.

**Physics-Grounded Learning and Machine-Learning Interatomic Potentials (MLIPs)**

HSMs are often evaluated in terms of equilibrium capacity and hydride formation stability, yet their practical performance is governed by dynamics.[43] Hydrogen diffuses through solids, redistributes among interstitial sites, interacts with defects, and drives the nucleation and evolution of hydride phases during cycling.[44] These processes are further influenced by volume changes and internal stresses, which can enhance transport or induce hysteresis, fracture, and degradation. **Hydrogen storage is therefore governed by evolving atomic and microstructural states rather than static structures.** Notably, hydrogen-containing systems often exhibit significant dynamical disorder, with some phases stabilized only at finite temperature, as reflected by imaginary phonon modes in static calculations; capturing such effects, including their impact on ionic mobility, remains a key challenge for both data representation and atomistic modeling.

This distinction has direct implications for atomistic modeling. DFT has been indispensable for evaluating hydrogen binding, migration barriers, and defect energetics, but is limited to small-scale, largely static simulations. *Ab initio* molecular dynamics (AIMD) extends this framework but remains restricted to short trajectories and cannot capture converged diffusion, phase boundaries, or long-time degradation. Mesoscale approaches describe phase-front propagation, chemo-mechanical coupling, and hysteresis, but depend on atomistic inputs that are often inconsistent and valid only over limited regimes.[45] A central challenge is therefore achieving DFT-level accuracy at experimentally relevant spatial and temporal scales.

**MLIPs provide a critical bridge (Figure 2a)**.[46] By learning potential-energy surfaces from DFT data, MLIPs enable simulation of hydrogen motion, structural rearrangement, and mechanical response across extended scales. They replace explicit electronic-structure calculations with surrogate models that predict energies, forces, and stresses from atomic environments.[47] This capability is particularly important for solid-state HSMs, where wide composition ranges, competing phases, and defect-rich environments require models that remain reliable across diverse configurations.[48] Active learning has become central to MLIP development, iteratively expanding the training domain through targeted DFT calculations.[49]

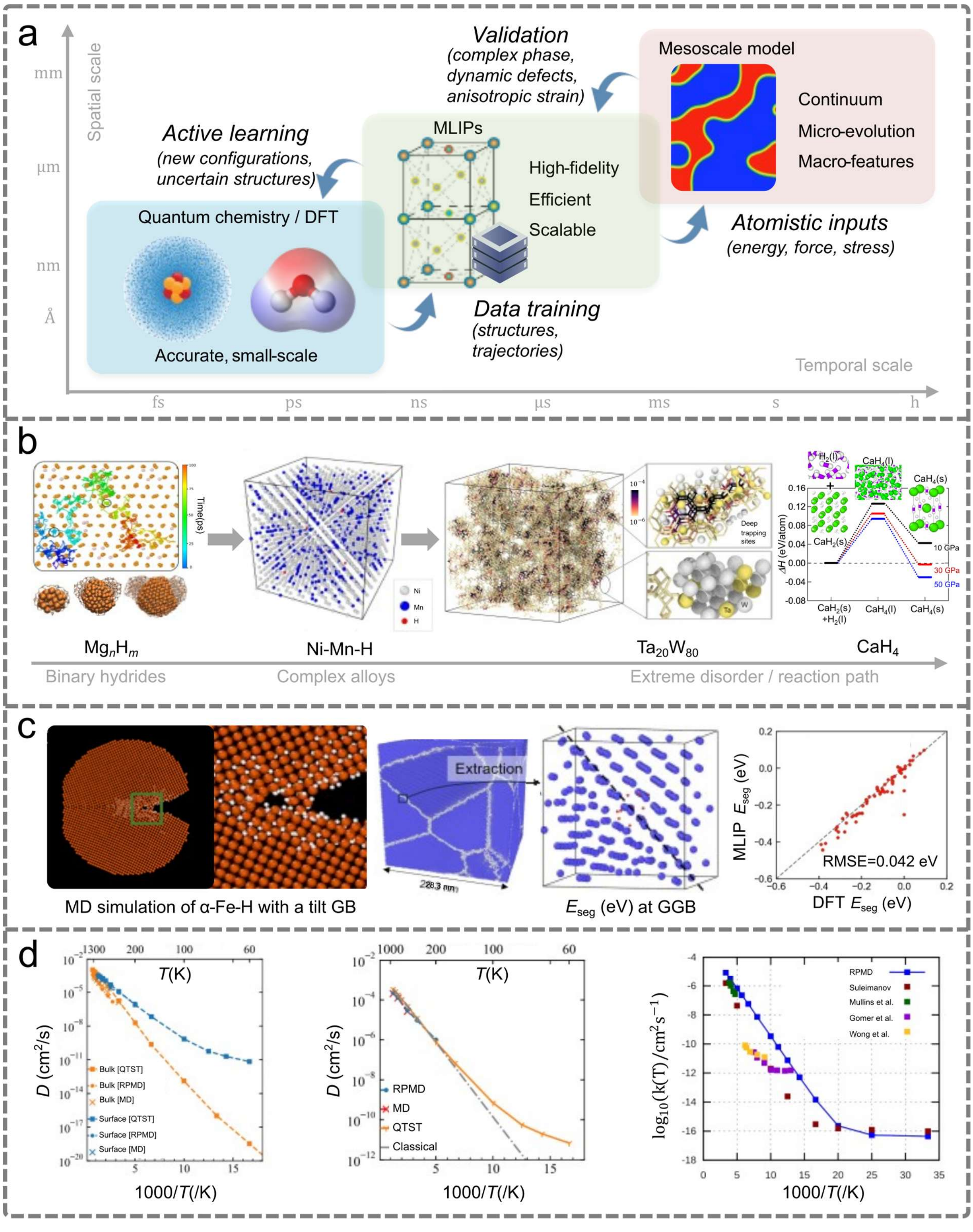

**Figure 2. Machine-learning interatomic potentials (MLIPs)-enabled multiscale modeling for hydrogen storage materials. (a)** Multi-scale modeling. MLIPs serve as a link bridging the spatiotemporal gap. Trained on quantum-fidelity but expensive data, MLIPs achieve density functional theory (DFT)-accuracy at speed of classical molecular dynamics (MD) simulations, and continuously self-refine *via* active learning with labeled uncertain configurations. MLIPs feed atomistic data into mesoscale models, which provide validation of complex phases, dynamic defects, and anisotropic strain, before supplying quantum-informed parameters into macroscopic applications. **(b)** Modeling of hydrogen diffusion across environments. **(left)** Time-progression trajectory (blue to red) of a hydrogen in $MgH_{0.0625}$ at 673 K, illustrating oscillations within interstitial sites.[51] Reproduced from Ref. 51 under the terms of the Creative Commons Attribution 4.0 International License (CC BY 4.0). Equilibrated $Mg_nH_m$ nanoclusters at 300 K, where orange spheres and bi-color sticks denote unbonded Mg and Mg-H coordination, respectively.[52] Reproduced with permission from Ref. 52. Copyright 2020 American Physical Society. **(center)** Representative atomic configuration of a Ni-25 at.% Mn alloy used in MD to predict hydrogen diffusion coefficients.[53] Reproduced from Ref. 53 under the terms of the Creative Commons Attribution 4.0 International License (CC BY 4.0). **(right)** Atomistic mechanisms of super-Arrhenius diffusion in non-equimolar $Ta_{20}W_{80}$ at 500 K; the visualization maps H visitation frequencies at tetrahedral interstitial sites, highlighting the local metallic environments of specific trapping and inaccessible sites.[54] In addition, MLIP-driven molecular dynamics simulations capture surface-melting-driven hydrogen absorption pathways during high-pressure hydrogenation for $CaH_4$, providing direct insight into reaction mechanisms beyond equilibrium diffusion.[58] Reproduced from Refs. 54 and 58, under the terms of the Creative Commons CC BY-NC license and the Creative Commons Attribution 4.0 International License (CC BY 4.0), respectively. **(c)** Atomistic insights into hydrogen segregation and microvoid evolution at grain boundaries (GBs). **(left)** Visualization of microvoid nucleation (green box) within an $\alpha$-Fe system containing a tilt (GBs) at a 1.856% hydrogen concentration.[59] Reproduced from Ref. 59, under the terms of the Creative Commons CC BY-NC license. **(right)** Representative calculation cell showing the initial positions of H atoms (red) relative to Fe atoms (blue) used to evaluate segregation energies ($E_{seg}$) at general grain boundaries (GGB). $E_{seg}$ for 85 distinct sites at GGB shows high agreement between MLIP (vertical axis) and DFT (horizontal).[60] Reproduced from Ref. 60, under the terms of the Creative Commons Attribution 4.0 International License (CC BY 4.0). **(d)** Temperature($T$)-dependence and quantum effects of hydrogen diffusion on Pd(111) surface and bulk. **(left)** Pd(111) surface diffusion is faster than bulk Pd at low $T$. This disparity arises from environment-specific nuclear quantum effects and competing zero-point energy (ZPE) corrections in surface and bulk. The data is evaluated using quantum transition state theory (QTST), ring polymer molecular dynamics (RPMD), and classical MD.[63] **(center)** Surface diffusion $D$ on the Pd(111) evaluated using QTST, ring polymer molecular dynamics (RPMD), classical MD, and classical transition state theory. The classical method exhibits Arrhenius behavior (a straight line). In contrast, the path-integral simulations (QTST, RPMD) deviate from this straight line below 200 K, showing a clear non-Arrhenius behavior.[63] Reproduced from Ref. 63, under the terms of the Creative Commons Attribution 4.0 International License (CC BY 4.0). **(right)** An MLIP model coupled with RPMD and kinetic Monte Carlo (kMC) resolved $T$-dependent hydrogen diffusion coefficients on Ni(100) and literature comparison.[64] Reproduced with permission from Ref. 64. Copyright 2020 American Institute of Physics.

**Foundation models have further expanded this capability.**[50] These models aim for transferability across chemistries, reducing the need for system-specific training. However, hydrogen storage systems (characterized by defects, multiphase behavior, and anisotropy) remain challenging and often require system-specific validation.

A central application is hydrogen diffusion, a primary kinetic bottleneck. In magnesium hydrides, MLIP-based simulations extend diffusion modeling beyond static barriers and short AIMD trajectories.[51] These approaches reveal nanoscale transport and structural evolution inaccessible to conventional methods (**Figure 2b**, left).[52] In high-entropy alloys, MLIPs capture environment-dependent diffusion behavior arising from chemical disorder (**Figure 2b**, center).[53] Coupled with neural-network-driven kinetic Monte Carlo, they further reveal super-Arrhenius diffusion in systems such as MoNbTaW (**Figure 2b**, right).[54]

Beyond transport, MLIPs enable modeling of phase transformations and hydrogen ordering.[55] In $TiCr_2$ Laves phases, actively learned MLIPs combined with DFT and Monte Carlo methods predict hydrogen ordering and thermodynamics.[56] Multiphase MLIPs spanning HCP, BCC, and FCC titanium hydrides demonstrate that accurate diffusion modeling requires broad phase coverage.[57] MLIP-driven molecular dynamics has further revealed complex hydrogenation pathways, including surface-melting-driven absorption and high-pressure superhydride formation (**Figure 2b**, right).[58]

Realistic hydrogen storage behavior also depends critically on defects, surfaces, and interfaces. Defects trap hydrogen, surfaces might hinder and interfaces redirect transport, and cyclic lattice expansion generates stresses that influence kinetics and durability. MLIP-based simulations capture these coupled effects, including hydrogen embrittlement and grain-boundary behavior (**Figure 2c**).[59, 60] These mechanisms are central to long-term cycling performance,[61] particularly in heterogeneous systems such as reactive hydride composites and catalyzed complex hydrides, where interfaces further increase the dimensionality of the design space. Complementary experimental validation is therefore essential: neutron scattering techniques, with their unique sensitivity to hydrogen, enable direct observation of site occupancy, diffusion, and phase evolution, providing key benchmarks for atomistic simulations.

Hydrogen further challenges classical descriptions due to nuclear quantum effects. Zero-point motion and tunneling influence diffusion barriers and Arrhenius prefactors, as well as isotope dependence (**Figure 2d**). MLIPs enable tractable path-integral simulations,[62] capturing phenomena such as non-Arrhenius quantum diffusion on Pd(111) and H diffusion on Ni surfaces.[63, 64]

Despite these advances, current MLIPs remain limited in scope and thermodynamic consistency. Their accuracy often degrades as hydrogen concentration changes, defects accumulate, or multiphase transformations occur—conditions central to real operation.[57] Hydrogen storage is governed by free-energy landscapes arising from coupled electronic, structural, and magnetic contributions.[65] Capturing these landscapes, rather than local forces alone, is essential for predicting experimentally observable behavior.

Progress therefore requires integrating MLIPs with explicit thermodynamic formalisms, including thermodynamic integration, free-energy methods, grand canonical ensembles, and CALPHAD-based approaches to predict plateau pressures, phase coexistence, and hysteresis.[66] Multiscale coupling provides a complementary pathway.[67] MLIP-derived quantities (e.g., diffusion coefficients, energetics, and interfacial properties) can parameterize mesoscale models, enabling simulation of nucleation, coarsening, and phase evolution.[68] Increasingly, these properties are recognized as size- and state-dependent, evolving with local structure and composition.[69, 70]

Importantly, most current case studies rely heavily on theoretical calculations. While these provide critical mechanistic insight, they remain limited by approximations in electronic structure methods and by incomplete representation of experimental complexity. A key future direction is therefore the integration of active learning with high-quality experimental characterization (e.g., atomic-scale structural analysis and *in situ* hydrogenation/dehydrogenation measurements) to iteratively calibrate simulations and refine underlying physical models. Such integration offers a pathway not only to improved predictive accuracy but also to the reconstruction of theoretical understanding based on experimentally validated behavior.

The next generation of multiscale frameworks will therefore combine DFT-level accuracy, MLIP-enabled atomistic dynamics, and mesoscale modeling into a unified, physics-grounded hierarchy capable of capturing microstructural evolution and long-term degradation under realistic conditions.

Beyond predictive fidelity, design-oriented modeling also benefits from explainability. For solid-state HSMs, it is often insufficient to reproduce capacity or diffusion alone; models must relate predictions to chemically meaningful local environments to guide experimental design. Structure-aware approaches provide one route, enabling site-resolved interpretation of model predictions within extended frameworks.

MLIPs have transformed hydrogen storage modeling from static descriptions of binding and barriers to dynamic descriptions of transport, phase transformation, and mechanics. **The next challenge is their integration into a thermodynamically coherent, multiscale framework that links atomic-scale processes to macroscopic performance, while enabling inverse design of solid-state HSMs.**

## Generative and Large Language Models

**The discovery of high-performance HSMs can be framed as a constrained inverse design problem.** Candidate materials must satisfy multiple competing criteria, including capacity, equilibrium pressure, kinetics, and cycling stability, with the requirement of falling within a narrow thermodynamic window representing a key—but not exclusive—constraint for reversible operation under practical conditions. Importantly, both enthalpy and entropy contribute to this constraint; in metal hydrides, these quantities are often coupled, such that reducing thermodynamic stability does not necessarily translate into improved operating conditions. Traditional high-throughput screening struggles to meet these requirements, as it is restricted to existing computational and experimental datasets.[43]

To address this limitation, recent approaches explore a "dual-stream" AI framework combining generative models (GMs) and LLMs. GMs act as structural generators, proposing candidate materials in continuous configuration space, while LLMs extract empirical knowledge and constraints from unstructured literature.

By learning structural patterns of known materials, GMs generate candidate structures from latent representations, expanding exploration beyond enumerated databases. This capability has enabled the identification of previously unreported structures, including new Mg-H phases and property-conditioned crystal generation using diffusion-based models such as MatterGen (**Figure 3a**).[71] However, because these models remain constrained by their training distributions and representations, they often explore variations of known structural motifs. More critically, while recent approaches incorporate property conditioning or post hoc filtering, the enforcement of physical validity, synthesizability, and hydrogen-storage-specific constraints during generation remains limited, which can lead to metastable or non-viable candidates, particularly in large combinatorial spaces such as nanoporous materials.[72]

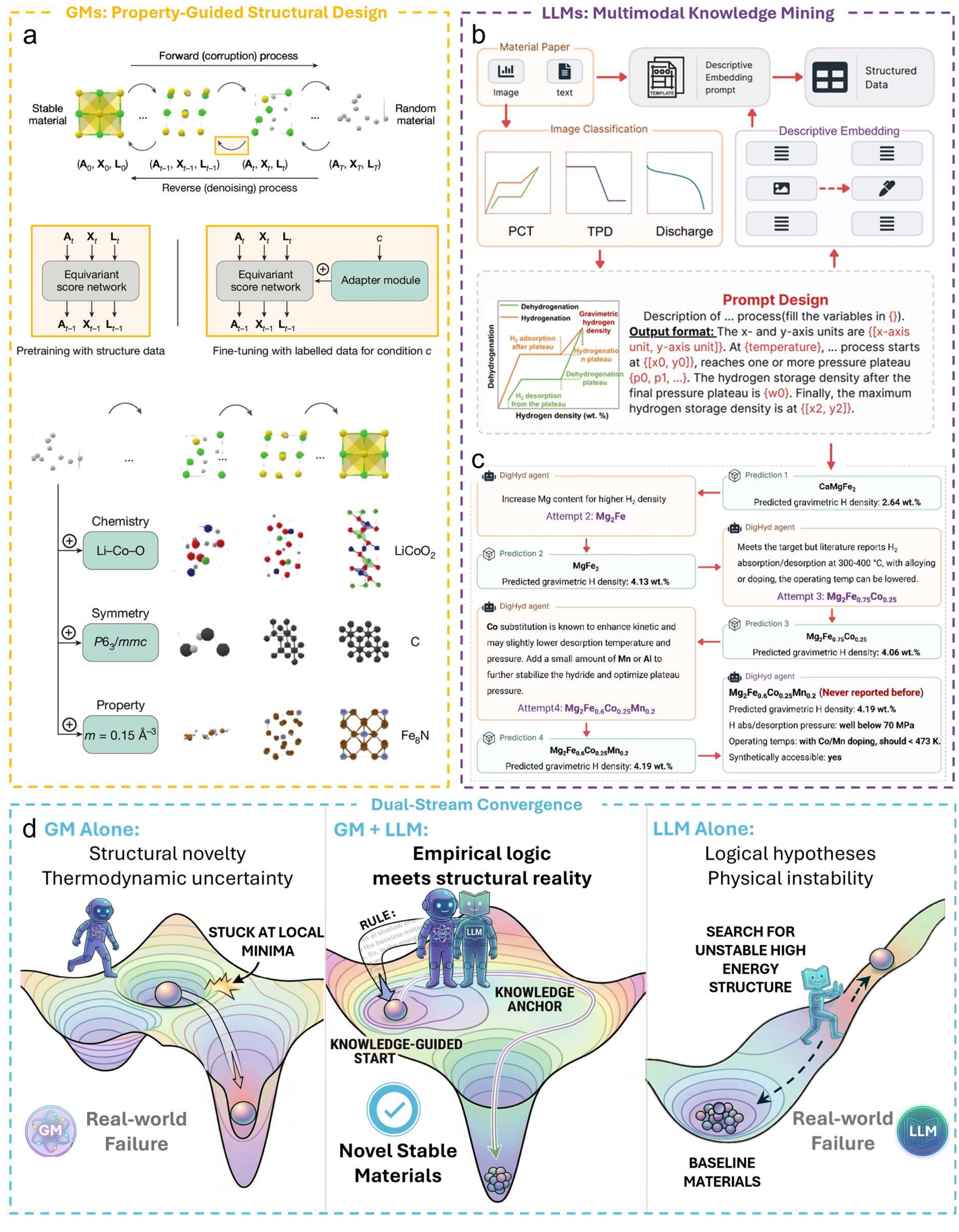

**Figure 3. The dual-stream generative model (GM)-large language model (LLM) architecture and agent-driven iterative inverse design for the constrained discovery of high-performance hydrogen storage materials. (a)** GMs: Property-Guided Structural Design. Generative models (e.g., MatterGen) navigate the continuous latent space to propose novel atomic configurations conditioned on user-defined metrics, such as specific chemistry, crystal symmetry, and targeted physical properties; in the context of hydrogen storage, this framework could be extended to incorporate constraints such as the $\Delta H$ window.[71] Reproduced from Ref. 71 under the terms of the Creative Commons Attribution 4.0 International License (CC BY 4.0). **(b)** LLMs: multimodal knowledge mining. The descriptive interpretation of visual expression workflow processes unstructured scientific literature, translating complex multimodal data into structured databases containing empirical synthesis constraints and thermodynamic relations.[35] Reproduced from Ref. 35, under the terms of the Creative Commons CC BY-NC license. **(c)** LLM agents: iterative inverse design. Building upon the mined historical database, the *DigHyd* agent executes an autonomous design-prediction-optimization loop.[35] The agent proposes initial candidates, which are dynamically evaluated by a pretrained ML regressor. The agent then iteratively reasons and refines the composition (e.g., evolving from $CaMgFe_2$ to a novel $Mg_2Fe_{0.6}Co_{0.25}Mn_{0.2}$ alloy) within minutes until researcher-defined thermodynamic targets are met. Reproduced from Ref. 35, under the terms of the Creative Commons CC BY-NC license. **(d)** Conceptual illustration of dual-stream convergence between GMs and LLMs for HSMs discovery. The integrated GM+LLM framework combines structural exploration with knowledge-based constraints, enabling knowledge-guided navigation of the energy landscape toward physically viable and synthetically accessible stable materials.

In parallel, LLMs are transforming how materials knowledge is utilized. Automated workflows such as DIVE convert graphical data (e.g., pressure-composition-temperature curves) into structured datasets, enabling large-scale database construction (**Figure 3b**).[35] Beyond data extraction, LLMs are increasingly integrated into active learning workflows, where they assist in proposing and refining candidate materials (either in combination with surrogate models or as direct decision-making agents), thereby accelerating iterative materials discovery (**Figure 3c**).[73, 74] However, LLM-generated hypotheses are not inherently constrained by physical laws and require external validation.

**These complementary limitations motivate integration. In a unified framework, LLMs provide empirically grounded constraints (e.g., composition limits, element selection, symmetry preferences, synthesis conditions, hydrogen occupancy ranges, and target property windows) while GMs generate explicit structural candidates that satisfy these inputs (Figure 3d).** Coupling this framework with physical validators (e.g., DFT or MLIP-based

simulations) enables thermodynamically consistent screening prior to experimental validation.[75, 76] Ultimately, embedding such architectures within closed-loop workflows linking design, simulation, and experiment offers a pathway toward autonomous, physics-aware discovery of solid-state HSMs.

**Closed-loop Discovery**

The discovery of HSMs is constrained by the interplay between high-dimensional compositional spaces and experimental variables, particularly in compositionally complex systems such as high-entropy alloys, where vast combinatorial spaces and disorder render exhaustive exploration infeasible. Unlike conventional equilibrium materials design, hydrogen storage performance depends not only on composition and structure but also on synthesis pathways, microstructural evolution, reaction mechanisms, and cycling behavior. **As a result, models trained on static datasets often fail to capture this dynamic, path-dependent behavior, motivating closed-loop discovery frameworks that integrate AI with automated experimentation.**

To enable such integration, experimental validation must match the throughput and consistency of computational screening, especially when exploring chemical spaces exceeding $10^6$ combinations. **SDLs address this need by integrating robotic synthesis, automated characterization, and adaptive monitoring into unified workflows, enabling continuous data generation and standardized measurement.**[77] Integrated platforms for HSMs are emerging, including efforts at Tohoku University under the Green Technologies of Excellence (GteX) program toward autonomous experimental workflows.

Within these frameworks, prediction, synthesis, characterization, and model refinement are iteratively linked (**Figure 4a**). ML models propose candidates, automated platforms synthesize materials, and standardized experiments (e.g., pressure–composition–temperature measurements and operando characterization) provide feedback. Although such data are routinely collected, they are often incompletely reported, limiting reuse. These data refine models, and similar closed-loop approaches have accelerated discovery in other domains, including catalysis and inorganic synthesis.[78] However, current systems remain largely tool-centric, with limited adaptability to uncertainty or evolving conditions.

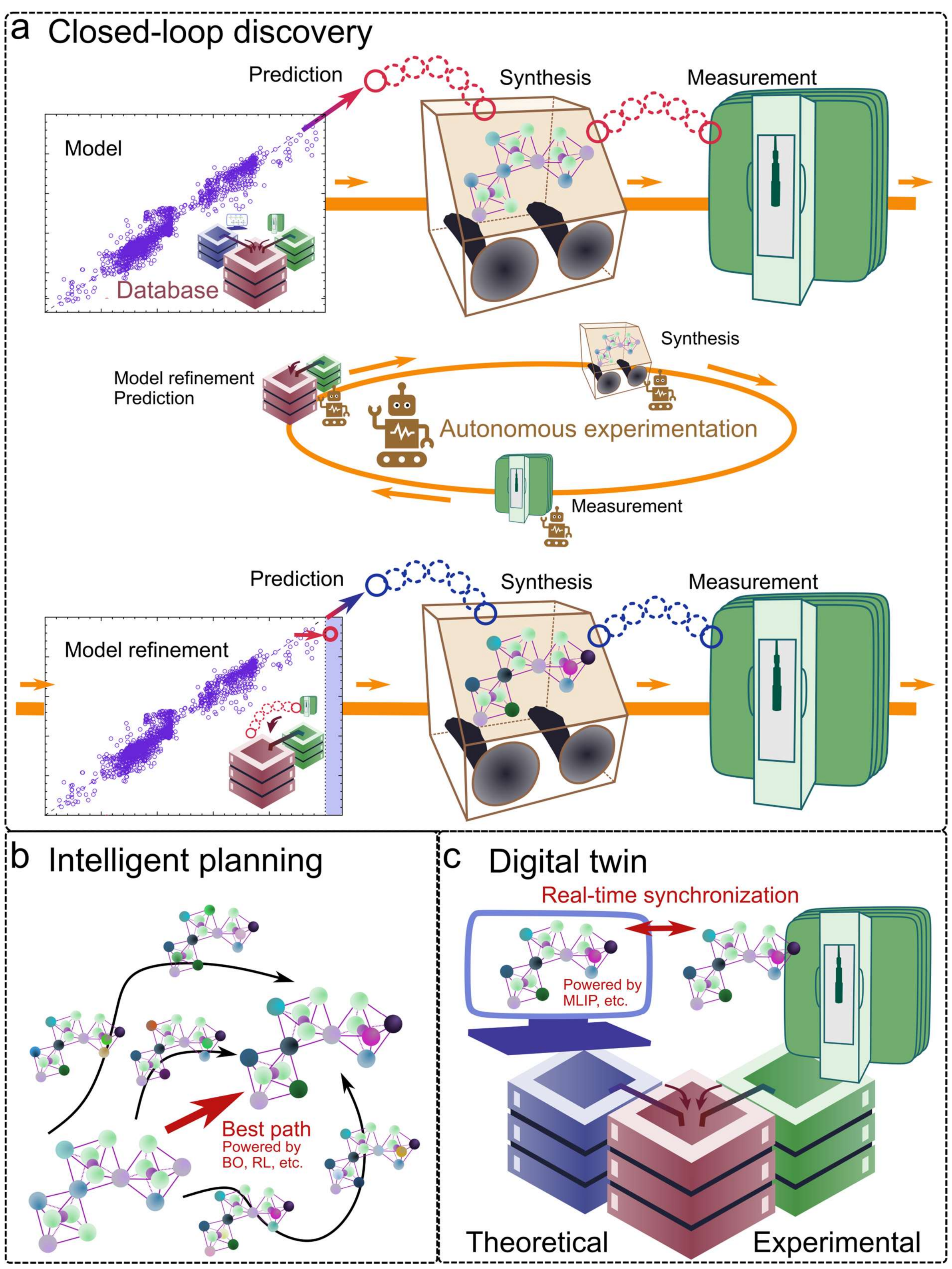

**Figure 4. Closed-loop discovery framework for hydrogen storage materials. (a)** Iterative closed-loop workflow for autonomous experimentation, integrating machine learning-driven

prediction, materials synthesis, and experimental measurement. Data generated from each iteration are fed back into the model, enabling continuous refinement and improved predictive accuracy. **(b)** Intelligent planning in high-dimensional compositional space. Adaptive strategies, such as Bayesian optimization (BO) and reinforcement learning (RL), guide the selection of candidate materials by balancing exploration and exploitation, thereby identifying optimal pathways toward improved hydrogen storage performance. **(c)** Concept of a digital twin for hydrogen storage systems. A dynamically updated virtual representation, powered by machine learning interatomic potentials (MLIPs) and related models, synchronizes computational predictions with experimental observations, enabling real-time feedback, model refinement, and physically consistent interpretation across theoretical and experimental domains.

The next stage therefore requires a shift from automation to intelligent planning (**Figure 4b**). Rather than selecting candidates solely based on predicted performance, closed-loop systems must incorporate uncertainty and prioritize experiments that improve model reliability. Bayesian optimization enables identification of regions where additional data most effectively reduce uncertainty,[78, 79] while reinforcement learning supports adaptive experiment selection by balancing exploration and exploitation.[80] This is particularly relevant for hydrogen storage, where datasets are uneven: properties such as capacity are relatively well characterized, whereas kinetics, cycling stability, and impurity sensitivity remain sparse. Effective frameworks must therefore allocate experimental effort toward reducing uncertainty in physically meaningful quantities, such as equilibrium pressure and degradation pathways.

Another critical challenge is the integration of heterogeneous data. Hydrogen storage experiments generate diverse outputs, including pressure-composition-temperature isotherms, kinetic measurements, structural characterization, and *operando* spectroscopy, each with different levels of abstraction and uncertainty. Closed-loop systems must therefore incorporate multimodal data fusion, enabling consistent interpretation across data types. This requirement reinforces the need for standardized data representations and robust data infrastructure.

A key direction is the development of a hydrogen storage "digital twin", a virtual representation that is dynamically updated and synchronizes computational predictions with real-time experimental observations (**Figure 4c**). At present, such approaches are most naturally applicable within a given class of materials, and extending them across chemically distinct systems remains an open challenge. By integrating MLIPs, thermodynamic models, and kinetic descriptions with continuously updated data, such systems enable co-evolution of simulation and experiment. Discrepancies between prediction and observation directly require model refinement, while real-time feedback enables adaptive experimental control and early detection of degradation mechanisms.

Realizing this vision requires addressing several challenges. Experimental data must be sufficiently standardized and reproducible for real-time integration, while MLIPs and surrogate models must accurately capture thermodynamic and kinetic behavior across diverse systems. In addition, computational and experimental components must be tightly coupled through robust interfaces to enable seamless feedback. These requirements highlight that closed-loop discovery is not solely an algorithmic problem, but a system-level challenge spanning data, models, and experimental platforms.

**Despite these challenges, closed-loop discovery represents a shift from static, dataset-driven research to dynamic, feedback-driven optimization.** For hydrogen storage, where performance is highly sensitive to experimental conditions and governed by coupled thermodynamic and kinetic processes, this approach provides a pathway toward efficient and physically informed exploration. Ultimately, success depends on integrating three key elements: reliable data infrastructure, physics-grounded models, and adaptive decision-making. When combined within a unified framework, these elements enable a transition toward systematic and autonomous optimization of HSMs.

## Conclusion

**This roadmap highlights that the discovery of high-performance HSMs is fundamentally a system-level challenge that cannot be addressed through isolated advances in data, modeling, or experimentation.** Progress is constrained less by the absence of algorithms than by fragmented data, limited physical fidelity across scales, insufficient attention to reproducibility, and weak integration between prediction and experiment. Addressing these challenges requires a shift toward a cohesive and interoperable research framework.

A key requirement is a thermodynamically consistent data infrastructure. Despite advances in computational and experimental databases, incomplete integration and inconsistent reporting continue to limit predictive reliability. Standardized ontologies, reporting protocols, and explicit links between datasets are therefore essential. At the modeling level, physics-grounded approaches (particularly MLIPs) bridge electronic structure accuracy and large-scale dynamics, but must be coupled with thermodynamic formalisms and multiscale frameworks. In parallel, integrating GMs and LLMs offers new opportunities for inverse design, provided predictions remain physically validated. **Ultimately, closed-loop discovery frameworks that integrate data, models, and automated experimentation provide a pathway toward autonomous, physically consistent discovery of next-generation solid-state HSMs.**

**Associated content**

**Author information**

**Acknowledgments**

This work was supported by the Green Technologies of Excellence (GteX) Program, Japan (Grant No. JPMJGX23H1). M. B. acknowledges support from the Project CH4.0 under the MUR program ''Dipartimenti di Eccellenza 2023-2027'' (CUP: D13C22003520001). K. S. acknowledges the ACS Petroleum Research Fund under Doctoral New Investigator Grant 68604-DNI6. J. P. acknowledges the University at Buffalo for start-up funds to support this work.

# ToC

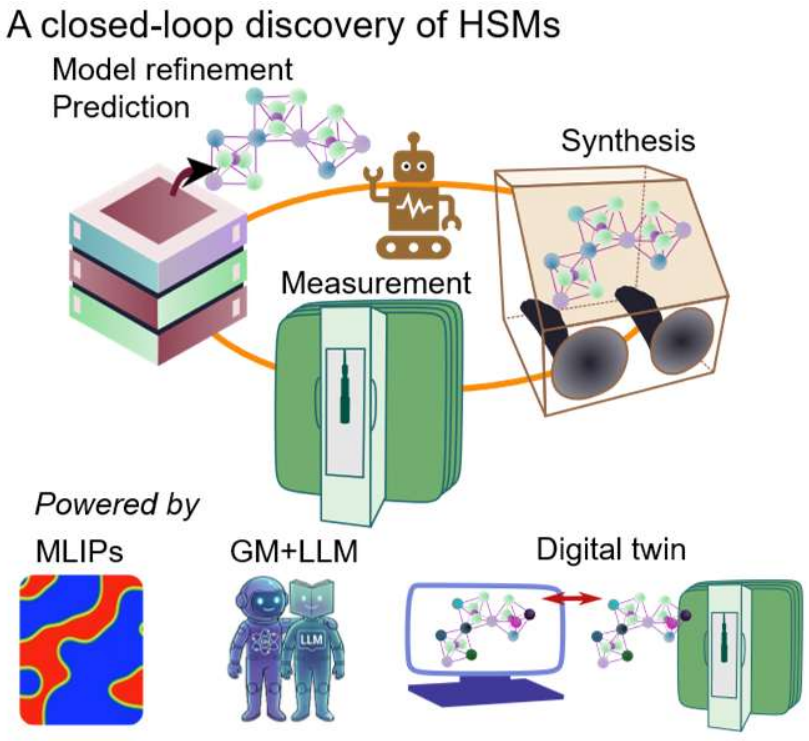
A closed-loop discovery of HSMs
Model refinement
Prediction
Synthesis
Measurement
Powered by
MLIPs
GM+LLM
Digital twin
LLM